\documentclass[aps,prl,showpacs,twocolumn]{revtex4}
\usepackage[latin1]{inputenc}
\usepackage{graphicx,amsmath,amsfonts,amssymb}
\usepackage{color}
\newcommand{\be}{\begin{equation}}
\newcommand{\ee}{\end{equation}}

\begin{document}

\title{Poisson's spot and Gouy phase}
\author{I. G. da Paz$^{1}$, Rodolfo Soldati $^{4}$, L. A. Cabral$^{2}$, J. G. G. de Oliveira Jr$^{3}$, Marcos Sampaio$^{4}$}

\affiliation{$^1$ Departamento de F\'{\i}sica, Universidade Federal
do Piau\'{\i}, Campus Ministro Petr\^{o}nio Portela, CEP 64049-550,
Teresina, PI, Brazil}

\affiliation{$^{2}$ Curso de F\'{\i}sica, Universidade Federal do
Tocantins, Caixa Postal 132, CEP 77804-970, Aragua\'{\i}na, TO,
Brazil}

\affiliation{$^{3}$ Departamento de Ci\^{e}ncias Exatas e
Tecnol\'{o}gicas, Universidade Estadual de Santa Cruz, Caixa Postal
45662-900, Ilh\'{e}us, BA, Brazil}

\affiliation{$^{4}$ Departamento de F\'{\i}sica, Instituto de
Ci\^{e}ncias Exatas, Universidade Federal de Minas Gerais, Caixa
Postal 702, CEP 30161-970, Belo Horizonte, Minas Gerais, Brazil}

\begin{abstract}
Recently there have been experimental results on Poisson spot matter
wave interferometry followed by theoretical models describing the
relative importance of the wave and particle behaviors for the
phenomenon. We propose an analytical theoretical model for the
Poisson's spot with matter waves based on Babinet principle in which
we use the results for a free propagation and single slit
diffraction. We take into account effects of loss of coherence and
finite detection area  using the propagator for a quantum particle
interacting with an environment. We observe that the matter wave
Gouy phase plays a role in the existence of the central peak and
thus corroborates the predominantly wavelike character of the
Poisson's spot.  Our model shows remarkable agreement with the
experimental data for deuterium ($D_{2}$) molecules.

\end{abstract}


\maketitle

\section{Introduction}
The wave nature of light which explains the Poisson's spot (``Tache
de Poisson-Fresnel-Arago") has an interesting history. In the
beginning of the $19^{th}$ century,  Fresnel submitted a paper on
the theory of diffraction supporting the wave nature of light for a
contest sponsored by the French Academy. Poisson, a member of the
judging committee, used Fresnel's theory to show the odd
prediction that a bright spot should appear behind a circular
obstacle. Arago, another member of the committee, thus observed the
spot experimentally. Fresnel won the competition, and the phenomenon
is known in history as Poisson's or Arago's spot \cite{Fresnel}.

Particle interferometry far-field diffraction behind a grating  and
near-field interference behind an opaque sphere or disk, namely the
observation of the Poisson's spot for matter waves provide
experimental evidence that matter can exhibit wave-particle duality.
Technology has greatly evolved since electron diffraction in the
1920s to interferometry in a grating with macromolecules like
fullerene in the 1999s \cite{Zeilinger}. Poisson spot has been
demonstrated by means of matter-waves  with electrons \cite{Komrska}
and  deuterium molecules \cite{Thomas1}. Some theoretical models
study the feasibility of the Poisson spot setup for fullerene
\cite{Thomas2} and gold clusters \cite{Juffmann}. The transverse
coherence of the matter wave beam is achieved for a source pinhole
sufficiently far away from the obstacle. Thus multi-path
interference leads to a bright spot at the center of the shadow
region behind the obstacle. From the experimental viewpoint
\cite{Juffmann,Thomas3}, it is believed that the diffraction pattern
is significantly affected by the dispersive interaction between the
matter-waves and the obstacle namely modifying the width and the
height of the central Poisson spot, invalidating the Fresnel zone
construction and Babinet principle. They argue that the spot could
appear in the case of classical particles passing the obstacle
following deflected trajectories due to the attractive force towards
the obstacle (van-der-Walls). In addition to that, we have the
unavoidable edge-corrugation of the disc. In \cite{Thomas3} it was
studied  the effect of Casimir-Polder/van der Waals (CP/vdW)
dispersion forces on Poisson spot diffraction at a dielectric sphere
which may obscure the distinction between particle and wave nature.
Obviously these effects blemish the distinction between the quantum
and the classical description for large enough interaction
strengths, such that the appearance of the spot in itself is not
exclusively due to wave-like behaviour of the particles.

Recently it was  shown that two fundamental but seemingly independent optical
phenomena, namely the Poisson spot and the orbital angular momentum
(OAM) of light, can be well connected by a phase changing. It was
demonstrated that spiral phase modulation can be added to the optical
tool to effectively shape the diffraction of light which may have
potential applications in the field of optical manipulations
\cite{Zhang}.

In 1890 L. G. Gouy observed the effect of a phase shift in light
optics that further was named Gouy phase \cite{gouy1,gouy2}. The
physical origin of this phase attracted the attention of several
researchers as can be seen in the works
\cite{Visser1,simon1993,feng2001,yang,boyd,hariharan,feng98, Pang}.
As is known today, the Gouy phase shift appears in any kind of wave
that is submitted to transverse spatial confinement, either by
focusing or by diffraction through small apertures. When a wave is
focused \cite{feng2001}, the Gouy phase shift is associated to the
propagation from $-\infty$ to $+\infty$ and is equal to $\pi/2$ for
cylindrical waves (line focus), and $\pi$ for spherical waves (point
focus). In the case of diffraction by a slit it was shown that the
Gouy phase shift is $\pi/4$ \cite{Paz4}. The Gouy phase shift has
been observed in water waves \cite{chauvat}, acoustic \cite{holme},
surface plasmon-polariton \cite{zhu}, phonon-polariton \cite{feurer}
pulses, and recently in matter waves \cite{cond,elec2,elec1}. As
some examples of applications of Gouy phase we can mention: the Gouy
phase has to be taken into account to determine the resonant
frequencies in laser cavities \cite{siegman} or the phase matching
in high-order harmonic generation (HHG) \cite{Balcou} and to
describe  the spatial variation of the carrier envelope phase of
ultrashort pulses in a laser focus \cite{Lindner}. Also, it plays
important role in the evolution of optical vortex beams \cite{Allen}
as well as electron beams which acquire an additional Gouy phase
dependent on the absolute value of the orbital angular momentum
\cite{elec2}.

In the coherent matter wave context the Gouy phase has been explored
in \cite{Paz4,Paz1,Paz2,Ducharme}. Experimental realizations were
made in different systems such as Bose-Einstein condensates
\cite{cond}, electron vortex beams \cite{elec2} and astigmatic
electron matter waves using in-line holography \cite{elec1}. Matter
wave Gouy phases have interesting applications, for instance, they
serve as mode converters important in quantum information
\cite{Paz1}, in the development of singular electron optics
\cite{elec1} and in the study of non-classical (looping) paths in interference
experiments \cite{Paz3}.

It is the main purpose of this contribution to perform a complete
analytical calculation of partially coherent matter-wave Poisson's
spot due to an unidimensional obstacle. We also define a generalized
expression for the Gouy phase for partially coherent matter waves
and study the effect of this phase in the Poisson's spot intensity.
The shape of the diffraction pattern on the screen can be computed
using Babinet principle \cite{Wolf}: the superposition
principle implies that the wave amplitudes behind a slit of certain
length, $\psi_{slit}$ and behind the corresponding obstacle of the
same length, $\psi_{obst}$, must add up such that $\psi_{slit} +
\psi_{obst} = 1$. In order to keep track of all important phases
such as Gouy phase and display fully analytical results we consider
a gaussian-shaped slit (obstacle). In this way we may compare with
experimental results and clearly distinguish wave interference from
mutually induced dipoles which give rise to van der Waals-type
attracting forces on the particle towards the obstacle as well as
imperfections of the blocking object. In order to incorporate loss of coherence in our model,
we obtain the reduced density matrix of the particles evolving effectively and
autonomously according to a ``Boltzmann-type" master equation. The
effect of the environment is summarized by ``a collision term" in
the propagator which takes into account the decoherence, that is to
say, the damping of off-diagonal terms of the density matrix in
position representation just as in \cite{Viale}. The Gouy phase for
partially coherent light wave was treated in  \cite{Visser}
which define a generalized expression for the Gouy phase in terms of
the cross-spectral density. For a model of matter waves with loss of
coherence we do not have an expression for the Gouy phase. However,
since the cross-spectral density and density matrix have analogous
meaning, in this contribution we follow the treatment adopted in
\cite{Visser} and  define the Gouy phase as the phase of the
density matrix.

The article is organized as follows: in section II we use the
Babinet principle to obtain analytical expression for the Poisson
spot with coherent matter waves. In section III we obtain analytical
expression for the Poisson's spot with partially coherent matter
waves and define a generalized expression for the Gouy phase.  These
results are used in section IV  to analyze the existing experimental
data. We draw our concluding remarks in section V.

\section{Babinet principle: Poisson spot and Gouy phase}

In this section we model the Poisson spot problem using the Babinet
principle and show that the intensity at the detector depends
on the Gouy phase and plays an important role particularly at the central peak.

For the sake of simplicity we will treat with a coherent model in
order to demonstrate the action of the Babinet principle as well as
the contribution of the Gouy phase for the intensity. We shall
obtain simple analytical expression for the Poisson intensity which
enables us to distinguish the role played by each phase. A source of
particles positioned behind an opaque disc of radius $\beta$ emits
particles one-by-one and a detector browses over a screen of
detection. It is a good approximation as we shall see to suppose an
one-dimensional model in which quantum effects are manifested only
in the $x$-direction as depicted in Fig.1 by a red line along a
diameter of the disc.

\begin{figure}[htp]
\includegraphics[width=7.0 cm]{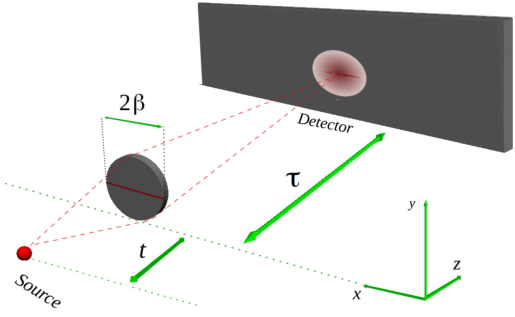}
\caption{Sketch of the Poisson spot problem. A source of particles
positioned behind of an opaque disc of radius $\beta$ send particles
one-by-one and a bright spot is observed by a detector in a screen
of detection. The red line along one diameter of the disc is used to
illustrate the treatment of the disc as a one dimensional
problem.}\label{Figure1}
\end{figure}

The propagation through the obstacle can be obtained by the Babinet
principle which enables us write
$\psi_{obst}(x,t,\tau)=\psi_{free}(x,t+\tau)-\psi_{slit}(x,t,\tau)$.
Here, $\psi_{obst}(x,t,\tau)$ stands for the wave function
describing the propagation through the obstacle,
$\psi_{free}(x,t+\tau)$ the wave function for free propagation and
$\psi_{slit}(x,t,\tau)$ the wave function characterizing propagation
through a single slit. To calculate the corresponding wave
functions, we consider that a coherent Gaussian wavepacket of
initial transverse width $\sigma_{0}$ is produced at the source and
propagates during a time $t$ before arriving at a single slit with
Gaussian aperture from which the Gaussian wavepacket propagates.
After crossing the slit the wavepacket propagates during a time
$\tau$ before arriving at detector in the detection screen. The
superposition of the wavepackets that propagate free and through the
slit gives rise to a interference pattern as a function of the
transverse coordinate $x$. Quantum effects are realized only in
$x$-direction as we consider that the energy associated with the
momentum of the particles in the $z$-direction is very high such
that the momentum component $p_{z}$ is sharply defined, i.e.,
$\Delta p_{z}\ll p_{z}$. Then we can consider a classical movement
in this direction with velocity $v_{z}$. Because the propagation is
free, the $x$, $y$ and $z$ dimensions decouple for a given
longitudinal location and thus we may write $z=v_{z}t$. Because
$v_{z}$ is assumed to be a well defined velocity we can neglect
statistical fluctuations in the time of flight, i.e., $\Delta t\ll
t$. Such approximation leaves the Schr\"{o}dinger equation analogous
to the optical paraxial Helmholtz equation \cite{Viale, Berman}.

The wave functions at the screen of detection are given by

\begin{equation}
\psi_{free}(x,t+\tau)=\int dx_{0} K_{t}(x,t+\tau;
x_{0},0)\psi_{0}(x_{0}),
\end{equation}

and

\begin{eqnarray}
\psi_{slit}(x,t,\tau)&=&\int  \int dx_{j} dx_{0} K_{\tau}(x,t+\tau;
x_{j},t)
 F(x_{j})\nonumber\\
 &\times&K_{t}(x_{j},t; x_{0},0)\psi_{0}(x_{0}),
\end{eqnarray}

with
\begin{equation}
K(x_{j},t_{j};x_{0},t_0)=\sqrt{\frac{m}{2\pi i\hbar
(t_{j}-t_{0})}}\exp\left[\frac{im(x_{j}-x_{0})^{2}}{2\hbar
(t_{j}-t_0)}\right],
\end{equation}

\begin{equation}
F(x_{j})=\exp\left[-\frac{(x_{j})^{2}}{2\beta^{2}}\right],
\end{equation}
and
\begin{equation}
\psi_{0}(x_{0})=\frac{1}{\sqrt{\sigma_{0}\sqrt{\pi}}}\exp\left(-\frac{x_{0}^{2}}{2\sigma_{0}^{2}}\right).
\end{equation}
The kernels $K_{t}(x_{j},t;x_{0},0)$ and
$K_{\tau}(x,t+\tau;x_{j},t)$ are the free propagators for the
particle, the function $F(x_{j})$ describes the slit transmission
function which is taken to be Gaussian of width $\beta$;
$\sigma_{0}$ is the effective width of the wavepacket emitted from
the source, $m$ is the mass of the particle, $t$ ($\tau$) is the
time of flight from the source (slit) to the slit (screen).

After some algebraic manipulations, we obtain

\begin{equation}
\psi_{free}(x,t+\tau)=\frac{1}{\sqrt{b\sqrt{\pi}}}\exp\left(-\frac{x^{2}}{2b^{2}}\right)\exp\left(\frac{imx^{2}}{2\hbar
r}+i\mu_{f}\right), \label{free}
\end{equation}

and

\begin{equation}
\psi_{slit}(x,t,\tau) = \frac{1}{\sqrt{B\sqrt{\pi}}}\exp
\left(-\frac{x^{2}}{2B^{2}}\right)\exp \left(\frac{imx^2}{2\hbar R}
+ i\mu_{s}\right), \label{slit}
\end{equation}

where

\begin{equation}
b(t+\tau)=\sigma_{0}\left[1+\left(\frac{t+\tau}{\tau_{0}}\right)^{2}\right]^{\frac{1}{2}},
\end{equation}

\begin{equation}
r(t+\tau)=(t+\tau)\left[1+\left(\frac{\tau_{0}}{t+\tau}\right)^{2}\right],
\end{equation}

\begin{equation}
\mu_{f}(t+\tau)=-\frac{1}{2}\arctan\left(\frac{t+\tau}{\tau_{0}}\right),
\end{equation}

\begin{equation}
B(t,\tau)
=\sqrt{\frac{\left(\frac{1}{\beta^{2}}+\frac{1}{b(t)^{2}}\right)^{2}+\frac{m^{2}}{\hbar^{2}}\left(\frac{1}{\tau}+\frac{1}{r(t)}\right)^{2}}
{\left(\frac{m}{\hbar\tau}\right)^{2}\left(\frac{1}{\beta^{2}}+\frac{1}{b(t)^{2}}\right)},}
\label{Bt}
\end{equation}

\begin{equation}
R(t,\tau)=\tau\frac{\left(\frac{1}{\beta^{2}}+\frac{1}{b(t)^{2}}\right)^{2}+\frac{m^{2}}{\hbar^{2}}\left(\frac{1}{\tau}+\frac{1}{r(t)}\right)^{2}}
{\left(\frac{1}{\beta^{2}}+\frac{1}{b(t)^{2}}\right)^{2}+\frac{t}{\sigma_{0}^{2}b(t)^{2}}\left(\frac{1}{\tau}+\frac{1}{r(t)}\right)},
\label{Rt}
\end{equation}

\begin{equation}
\mu_{s}(t,\tau)=-\frac{1}{2}\arctan\left[\frac{t+\tau\left(1+\frac{\sigma_{0}^{2}}{\beta^{2}}\right)}{\tau_{0}\left(1-\frac{t\tau\sigma_{0}^{2}}{\tau_{0}^{2}
\beta^{2}}\right)}\right] \label{Gouy1},
\end{equation}

and

\begin{equation}
\tau_{0}=\frac{m\sigma_{0}^{2}}{\hbar}.
\end{equation}

Here, $b(t+\tau)$, $r(t+\tau)$ and $\mu_{f}(t+\tau)$ are
respectively the beam width, the radius of curvature of the
wavefronts and Gouy phase for the free propagation during the total
time $t+\tau$. Moreover, $B(t,\tau)$, $R(t,\tau)$ and
$\mu_{s}(t,\tau)$ are respectively the beam width, the radius of
curvature of the wavefronts and Gouy phase for the propagation
through a single slit. $B(t,\tau)$ and $R(t,\tau)$ can be  written
in terms of $b(t)$ and $r(t)$, i.e, the beam width and the radius of
curvature of the wavefronts for the free evolution from the source
to the slit (or disc). The parameter
$\tau_{0}=m\sigma_{0}^{2}/\hbar$ is viewed as a characteristic time
for the ``aging" of the initial state \cite{solano}.

According to the Babinet principle the intensity at the screen of
detection is given by

\begin{eqnarray}
I(x,t,\tau)&=&|\psi_{obst}(x,t,\tau)|^{2}\nonumber\\
&=&\frac{1}{\sqrt{\pi}b}\exp\left(-\frac{x^{2}}{b^{2}}\right)+\frac{1}{\sqrt{\pi}B}\exp\left(-\frac{x^{2}}{B^{2}}\right)\nonumber\\
&-&\frac{2}{\sqrt{\pi b
B}}\exp\left[-\left(\frac{1}{2b^{2}}+\frac{1}{2B^{2}}\right)x^{2}\right]\nonumber\\
&\times&\cos\left[\frac{mx^{2}}{2\hbar}\left(\frac{1}{R}-\frac{1}{r}\right)+\mu(t,\tau)\right],
\label{I_Free}
\end{eqnarray}
where
\begin{eqnarray}
\mu(t,\tau)&=&\mu_{s}(t,\tau)-\mu_{f}(t+\tau)\nonumber\\
&=&
-\frac{1}{2}\arctan\bigg\{\frac{\tau[\tau_{0}^{2}+t(t+\tau)]}{\tau_{0}\tau^{2}+\frac{\beta^{2}}{\sigma_{0}^{2}}\tau[(t+\tau)^{2}+\tau_{0}^{2}]}\bigg\}
\label{gouy_c}
\end{eqnarray}
is the coherent Gouy phase difference. Therefore, from equation
(\ref{I_Free}) we clearly observe the Gouy phase effect on the
Poisson's  spot intensity. We illustrate such an effect in Fig. 2 by
plotting the normalized intensity $I$ for the parameters of
deuterium molecules of Ref. \cite{Thomas1}, i.e.,
$m=3.34\times10^{-27}\;\mathrm{kg}$, $\sigma_{0}=50\;\mathrm{\mu m}$
and $\beta=60\;\mathrm{\mu m}$. We consider the propagation times
$t=20\;\mathrm{ms}$ and $\tau=40\;\mathrm{ms}$. For solid line we
consider and for pointed line we do not consider the Gouy phase
effect. A pronounced peak at $x=0$ appears for the case in which we
consider the Gouy phase difference.

\begin{figure}[htp]
\centering
\includegraphics[width=5.0 cm]{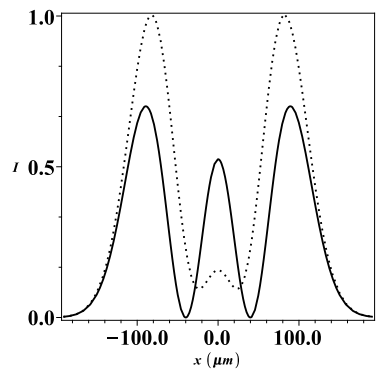}
\caption{Gouy phase effect on the Poisson spot for a coherent model.
Solid line we consider and for pointed line we do not consider the
Gouy phase effect on the normalized intensity $I$.}\label{Gouy}
\end{figure}

The dependence of the Poisson's spot intensity on the Gouy phase
obviously appears corroborates the wave nature of the Poisson's spot
since the Gouy phase is a wave property. The
simple model treated so far in this section does not take into account some
effects that a realist model to Poisson spot have to present. It would be
interesting to investigate if a more realistic
model for the Poisson spot can still be related to the Gouy phase.
It is the purpose of the next section to investigate such extension.

\section{A model with loss of coherence}

The result obtained in equation (\ref{I_Free}) for the Poisson's
spot intensity does not take account any loss of coherence. We shall
consider that the loss of coherence is produced from the obstacle to
the screen and therefore starting from time $t+\epsilon$ (with
$\epsilon\rightarrow0$ being the propagation time through the
obstacle) until the detection screen. Now, the evolution during the
time $\tau$ is given by the propagator for a quantum particle
interacting with an environment. In order to include such a loss of
coherence we follow the result obtained in Ref. \cite{Viale} and
write the Poisson spot intensity as

\begin{eqnarray}
I_{\ell}(x,t,\tau)&\equiv &\rho(x=x^{\prime},t,\tau)=N\int\int dx_{0}dx_{0}^{\prime}\nonumber\\
&\times&\exp\left\{\frac{im}{2\hbar
\tau}[x_{0}^{2}-x_{0}^{\prime2}+2x(x_{0}-x_{0}^{\prime})]\right\}\nonumber\\
&\times&\exp\left[-\frac{(x_{0}-x_{0}^{\prime})^{2}}{2\ell^{2}(\tau)}\right]
\tilde{\rho}(x_{0},x_{0}^{\prime},t), \label{int_l}
\end{eqnarray}
where

\begin{equation}
\tilde{\rho}(x_{0},x_{0}^{\prime},t)=\psi_{obst}(x_{0},t,\epsilon\rightarrow0)\psi_{obst}^{*}(x_{0}^{\prime},t,\epsilon\rightarrow0),
\end{equation}

and

\begin{eqnarray}
\ell(\tau)\equiv\frac{\ell_{0}}{\sqrt{1+\frac{2\Lambda
\tau}{3}}\ell_{0}^{2}}.
\end{eqnarray}
Here, $N$ is a normalization constant,
$\tilde{\rho}(x_{0},x_{0}^{\prime},t)$ is the density matrix in the
obstacle, $\tau$ is the propagation time from the obstacle to the
screen in which we have loss of coherence, $\ell(\tau)$ is the time
dependent coherence length and $\ell_{0}$ is the coherence length in
the obstacle which is the same of the source since we consider that
the propagation from the source to the obstacle is free, i.e.,
$\ell_{0}=\ell(t)$. The parameter $\Lambda$ encodes decohering
events such as scattering and photon emission and $\ell_{0}$ carries
incoherence effects of the source \cite{Viale}.

In order to obtain the density matrix in the obstacle we have to
take the limit when $\epsilon\rightarrow0$ in the parameters
$B(t,\epsilon)$, $R(t,\epsilon)$ and $\mu_{s}(t,\epsilon)$ of the
wavefunction given by equation (\ref{slit}).  After performing such
limits using the expressions (\ref{Bt}), (\ref{Rt}) and
(\ref{Gouy1}), we  obtain the following results
$\lim_{\epsilon\rightarrow0}\;B(t,\epsilon)=\sqrt{\frac{b^{2}(t)\beta^{2}}{\beta^{2}+b^{2}(t)}}$,
$\lim_{\epsilon\rightarrow0}\;R(t,\epsilon)=r(t)$ and
$\lim_{\epsilon\rightarrow0}\;\mu_{s}(t,\epsilon)=\mu_{f}(t)$.
Notice that only the parameter $B(t,\epsilon\rightarrow0)$ is
changed by the slit. Using the results above we obtain the density
matrix in the obstacle $\tilde{\rho}(x_{0},x_{0}^{\prime},t)$.

After performing the integration in equation (\ref{int_l}) and some
algebraic manipulation we obtain

\begin{eqnarray}
I_{\ell}(x,t,\tau)&=&\sqrt{\frac{\pi}{\eta}}\exp\left[-\frac{m^{2}x^{2}}{4\eta\hbar^{2}\tau^{2}}\right]\nonumber\\
&+&\sqrt{\frac{\pi}{\eta^{\prime}}}\exp\left[-\frac{m^{2}x^{2}}{4\eta^{\prime}\hbar^{2}\tau^{2}}\right]\nonumber\\
&-&\frac{2\sqrt{2\pi\hbar
\tau}}{\sqrt{\sqrt{C[1+(b(t)/\beta)^{2}]}}}\exp(-\alpha
x^{2})\nonumber\\
&\times&\cos(\delta x^{2}+\mu_{\ell}), \label{I_Decoh}
\end{eqnarray}

where

\begin{eqnarray}
\eta(t,\tau)&=&b^{2}(t)\bigg[\frac{1}{2b^{2}(t)}\left(\frac{1}{2b^{2}(t)}+\frac{1}{\ell^{2}}\right)\nonumber\\
&+&\left(\frac{m}{2\hbar r(t)}+\frac{m}{2\hbar
\tau}\right)^{2}\bigg],
\end{eqnarray}

\begin{eqnarray}
\eta^{\prime}(t,\tau)&=&\left(\frac{\beta^{2}b^{2}(t)}{\beta^{2}+b^{2}(t)}\right)\bigg[\left(\frac{m}{2\hbar
r(t)}+\frac{m}{2\hbar
\tau}\right)^{2}\nonumber\\
&+&\left(\frac{1}{2b^{2}(t)}+\frac{1}{2\beta^{2}}\right)\nonumber\\
&\times&\left(\frac{1}{2b^{2}(t)}+\frac{1}{2\beta^{2}}+\frac{1}{\ell^{2}}\right)\bigg],
\end{eqnarray}

\begin{eqnarray}
\alpha(t,\tau)&=&\frac{m^{2}}{C}\left(\frac{1}{b^{2}(t)}+\frac{1}{2\beta^{2}}\right)\nonumber\\
&\times&\bigg[\left(\frac{\beta^{2}+b^{2}(t)}{4\beta^{2}b^{2}(t)}\right)\left(\frac{1}{b^{2}(t)}+\frac{1}{\ell^{2}}\right)\nonumber\\
&+&\frac{1}{4\ell^{2}b^{2}(t)}+\left(\frac{m}{2\hbar
r(t)}+\frac{m}{2\hbar \tau}\right)^{2}\bigg],
\end{eqnarray}

\begin{equation}
\delta(t,\tau)=\frac{m^{3}}{4\hbar
\beta^{2}C}\left(\frac{1}{b^{2}(t)}+\frac{1}{2\beta^{2}}\right)\left(\frac{1}{r(t)}+\frac{1}{\tau}\right),
\end{equation}
and
\begin{eqnarray}
C(t,\tau)&=&4\hbar^{2}\tau^{2}\bigg\{\frac{m^{2}}{16\hbar^{2}
\beta^{4}}\left(\frac{1}{r(t)}+\frac{1}{\tau}\right)^{2}\nonumber\\
&+&\bigg[\left(\frac{\beta^{2}+b^{2}(t)}{4\beta^{2}b^{2}(t)}\right)\left(\frac{1}{b^{2}(t)}+\frac{1}{\ell^{2}}\right)\nonumber\\
&+&\frac{1}{4\ell^{2}b^{2}(t)}+\left(\frac{m}{2\hbar
r(t)}+\frac{m}{2\hbar \tau}\right)^{2}\bigg]^{2}\bigg\}.
\end{eqnarray}
The Poisson spot intensity given by equation (\ref{I_Decoh}) is the
main result of this paper. To our knowledge, an analytical expression incorporating such
effects for the Poisson's spot has not
been obtained. The result of equation (\ref{I_Decoh}) is
useful to define the Gouy phase for partially coherent matter waves
and explore the role of this phase in the Poisson spot.

\subsection{Generalized Gouy phase for partially coherent matter waves}

In Ref. \cite{Paz1} was shown that the matter waves Gouy phase is
related with the off diagonal elements of the covariance matrix
which indirectly enabled to extract the Gouy phase from the beam
width. It was shown that the experimental data for the diffraction
of fullerene molecules is quantitatively consistent with the
existence of a Gouy phase. Since the fullerene molecules have to be
treated as partially coherent matter waves in that work was
conjectured that the Gouy phase can be obtained by integrating the
inverse of the squared beam width, as is valid for coherent case.
Further, a complete definition for the Gouy phase for partially
coherent light waves was given in Ref. \cite{Visser}. In this work
we use the definition of Ref. \cite{Visser} to obtain the Gouy phase
for partially coherent matter waves as
\begin{equation}
\mu_{\ell}(t,\tau)=\arg[I_{\ell}(0,t,\tau)].
\end{equation}
We have a complete analogy with the generalized definition for the
Gouy phase of Ref. \cite{Visser}, since here
$I_{\ell}(0,t,\tau)\equiv\rho(0,t,\tau)$ is the density matrix in
the propagation axis $z$ which is similar to the cross-spectral
density and $(t,\tau)$ can be used to obtain two different positions
in the propagation axis since we are substituting the propagation
time by $z/v_{z}$. In the case of light waves the mechanism of loss
of coherence is attributed to the source incoherence whereas for
matter waves  such effects can be attributed both to the source
incoherence and environment decoherence. We calculate the Gouy phase
here and obtain the following result
\begin{equation}
\mu_{\ell}(t,\tau)=-\frac{1}{2}\arctan\left[\frac{r(t)+\tau}{a_{1}+a_{2}+\frac{r(t)\tau}{\tau_{0}}\left(1+\frac{2\beta^{2}}{b^{2}(t)}\right)\frac{\sigma_{0}^{2}}{\ell^{2}}}\right],
\label{gouy_l}
\end{equation}
where
\begin{equation}
a_{1}(t,\tau)=\left(\frac{\beta^{2}\tau_{0}}{\sigma_{0}^{2}r(t)\tau}\right)(r(t)+\tau)^{2},
\end{equation}
and
\begin{equation}
a_{2}(t,\tau)=\frac{r(t)\tau}{\tau_{0}}\left(1+\frac{\beta^{2}}{b^{2}(t)}\right)\left(\frac{\sigma_{0}}{b(t)}\right)^{2}.
\end{equation}
We can observe from equation (\ref{gouy_l}) that the Gouy phase is
dependent of the coherence length $\ell$. The same dependence was
discussed in Ref. \cite{Visser} for partially coherent light wave.
We can easily obtain that the result of equation (\ref{gouy_l})
reduces to that of equation (\ref{gouy_c}) for coherent matter waves
in the limit $\ell\rightarrow\infty$. On the other hand, in the
limit of completely non-coherent matter waves $\ell\rightarrow0$ we
have $\mu_{\ell}\rightarrow0$.

Therefore, just as in the coherent case the Poisson spot intensity is
changed by the Gouy phase. This can be clearly seen in the figure below. In
Fig. 3(a) we show the Gouy phase $\mu_{\ell}(t,\tau)$ as a function
of the propagation time $\tau$ for $t=20\;\mathrm{ms}$ and for the
data of the deuterium molecules. Solid line corresponds to
$\ell=1.0\;\mathrm{m}$ and pointed line corresponds to
$\ell=100\;\mathrm{\mu m}$. In Fig. 3(b) we show the normalized
intensity $I_{\ell}$ as a function of $x$ for $t=20\;\mathrm{ms}$
and $\tau=40\;\mathrm{ms}$ for the data of deuterium molecules.
Solid line we consider the effect of the phase $\mu_{\ell}$ and
pointed line we do not consider such effect.
\begin{figure}[htp]
\centering
\includegraphics[width=4.2 cm]{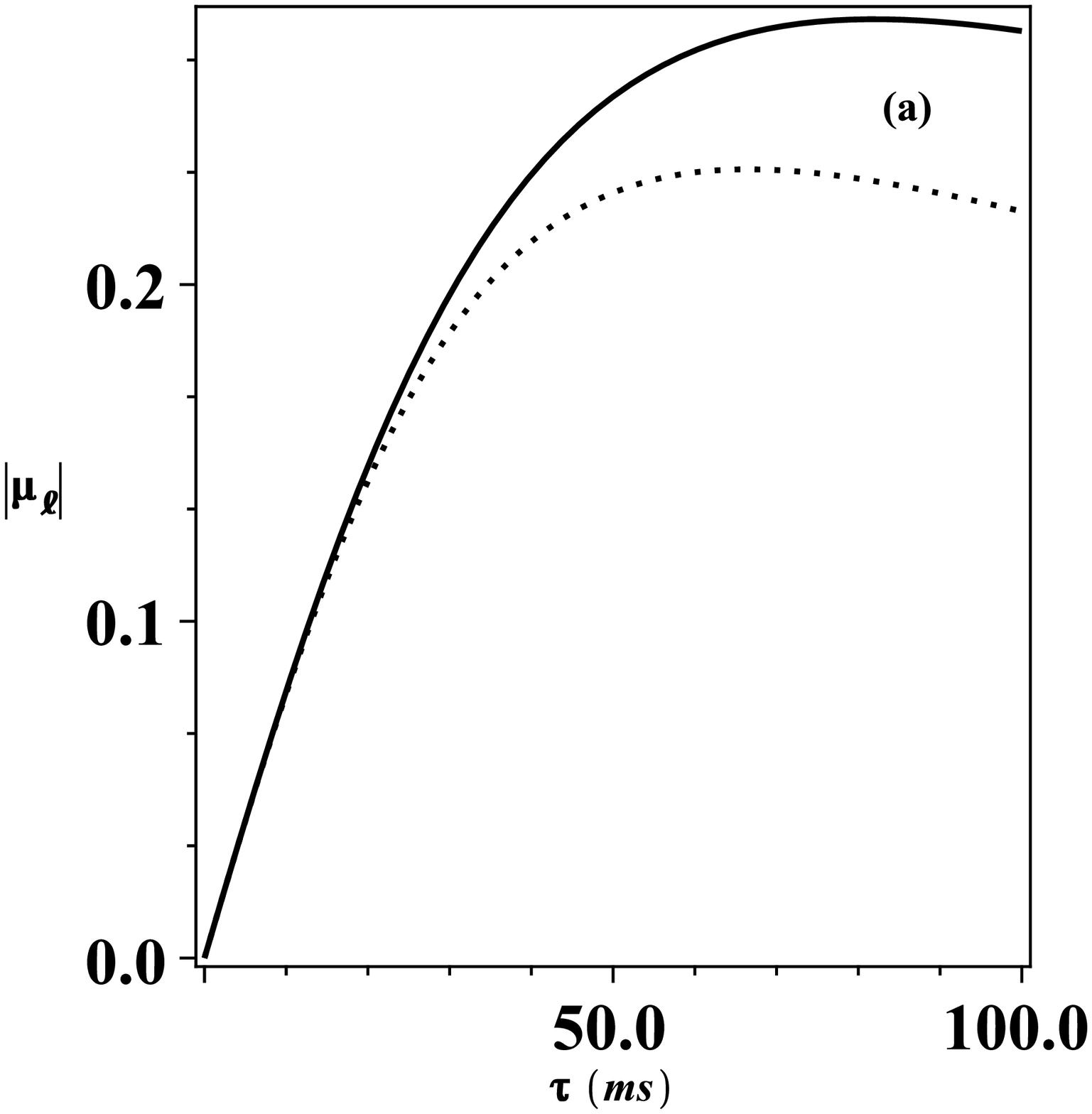}
\includegraphics[width=4.2 cm]{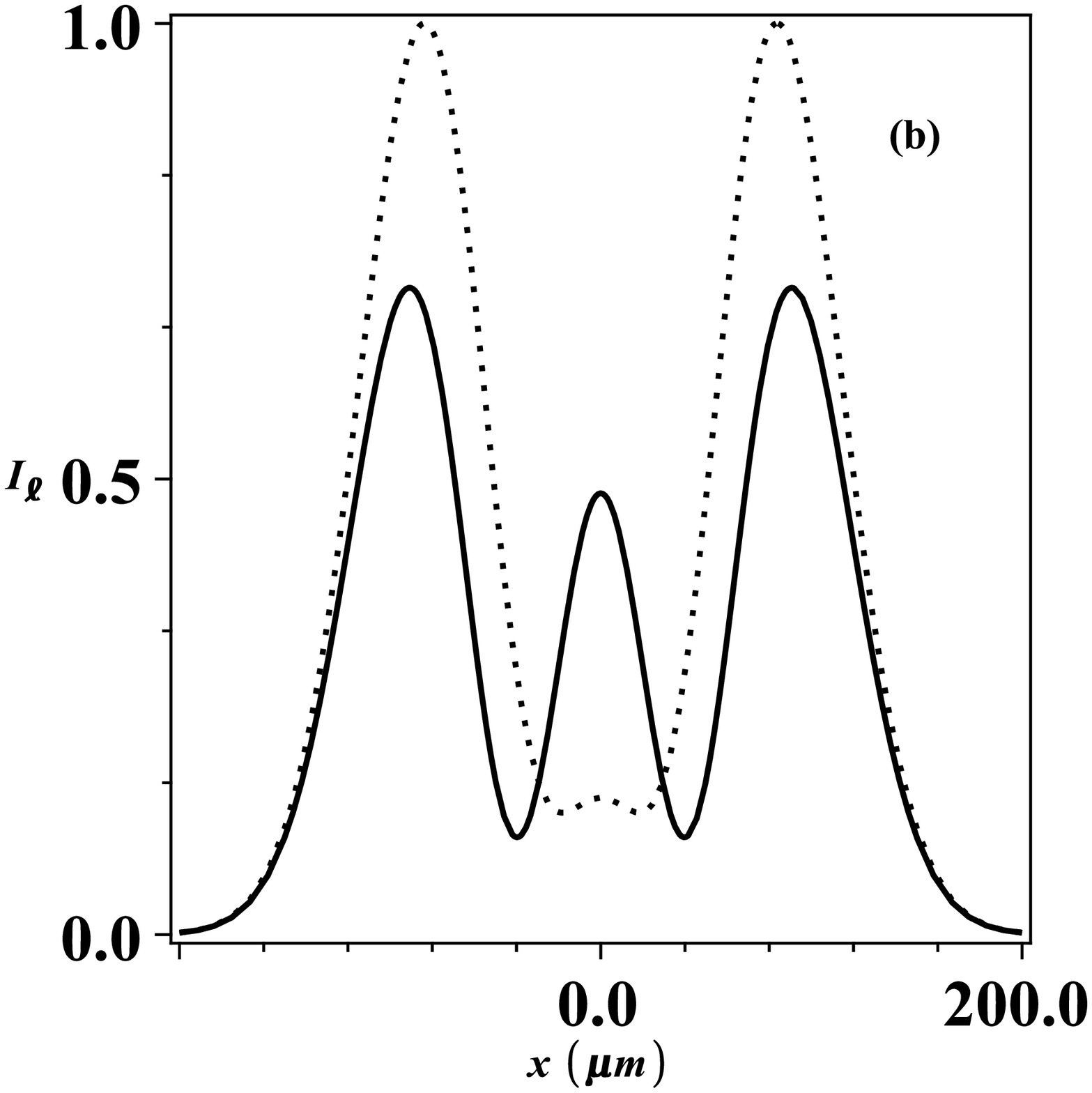}
\caption{(a) Gouy for partially coherent matter waves as a function
of $\tau$ for two different values of coherence length and
$t=20\;\mathrm{ms}$. Solid line corresponds to
$\ell=1.0\;\mathrm{m}$ and pointed line corresponds to
$\ell=100\;\mathrm{\mu m}$. (b) Normalized intensity $I_{\ell}$ as a
function of $x$ for $\ell=100\;\mathrm{\mu m}$, $t=20\;\mathrm{ms}$
and $\tau=40\;\mathrm{ms}$. Solid line we consider the effect of the
phase $\mu_{\ell}$ and pointed line we do not consider such
effect.}\label{gouy_dec}
\end{figure}

Fixing a set values of parameters we observe that the Gouy phase
decreases when the coherence length decreases. This is a novel
result in the context of matter waves. The visibility of the Poisson
spot tends to decrease as an effect of the loss of coherence. We can
observe this by comparing the intensity for the coherent and
partially coherent case, Fig. 2 and Fig. 3(b) respectively, which
show that the minimum intensity for the partially coherent case is
not zero. The partially coherent Gouy phase changes the intensity in
a such way that the intensity in the central peak is not observed if
one neglected this phase. The presence of the Gouy phase is a
signature of the wave behaviour. Thus, the relationship between
Poisson spot and Gouy phase for a model of partially coherent matter
waves with analytical results as obtained in this section is useful
to treat experimental data and to elucidate the wave behaviour of
the Poisson spot with matter waves. In order to test our results in
the next section we will analyse the experimental data for deuterium
molecules of Ref. \cite{Thomas1}.

\section{Analysis of existing experimental data}

In this section we compare our model with the experimental data for
deuterium molecules of Ref. \cite{Thomas1}. We take into account loss of
coherence and the finite detection area. The loss of coherence was
obtained in equation (\ref{I_Decoh}). To include the detector effect
we perform a convolution to obtain the effective intensity
\begin{equation}
I_{eff}(x,t,\tau)=\int^{\infty}_{-\infty}
I_{\ell}(x^{\prime},t,\tau)D(x-x^{\prime})dx^{\prime}.
\end{equation}
Considering a Gaussian profile to the detector aperture as
$D(x)=\exp\left(-x^{2}/2\sigma_{D}^{2}\right)$, where $\sigma_{D}$
is the detector width, the integral above is easily done.

In order to compare our model with experimental results previously
published in Ref. \cite{Thomas1} we relate our model and the one in
\cite{Thomas1} by $I_{eff}(x,l,\sigma_D)=a+b\;I(x,l,\sigma_D)$. The
parameters $a$ and $b$ are necessary to convert our results in units
($rate/s$) used in Ref. \cite{Thomas1}. The numerical calculations
obtained within these units are summarized in Table 1. The obtained
results are in good agreement with the experimental values, as we
can observe in Fig. 4.

\begin{table}[t]
\caption{\label{table1}Parameters of analytic model and numerical
results}
\begin{ruledtabular}
\begin{tabular}{ll}
 Coherence parameter & $\ell=0.3369\;\mathrm{\mu m}$ \\
Detector width & $\sigma_{D}=3.96\;\mathrm{\mu m}$ \\
Gaussian width & $\sigma_0=51\;\mathrm{\mu m}$\\
Disc aperture \footnote{The physical parameters are compatible with Ref. \cite{Thomas1}}& $\beta=60\;\mathrm{\mu m}$\\
Time before disc & $t=1.4\;\mathrm{ms}$\\
Time after disc & $\tau=0.606\;\mathrm{ms}$\\
Partially coherent fit [PC] $^{b}$ & $a=29829.11, \; b=-348.71$\\
Detector convolution fit [DC] \footnote{See the corresponding curve on Fig. 4 and experimental data [EX] from Ref. \cite{Thomas1}} & $a=40465.09,\; b=-466.29$  \\
%Experimental data [EX] & (See Fig.4) of Ref. \cite{Thomas1}\\
Gouy Phase [PC] & $\mu_\ell = 0.00060097028\;\mathrm{rad}$  \\
Gouy Phase [DC]& $\mu_\ell = 0.00069360626\;\mathrm{rad}$\\
\end{tabular}
\end{ruledtabular}
\end{table}

\begin{figure}[htp]
\includegraphics[width=6.0 cm]{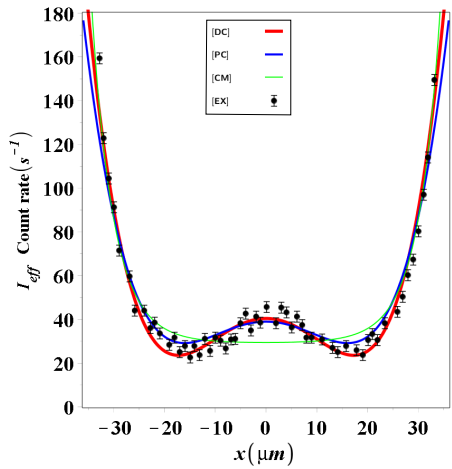}
\caption{Comparison of experimental data Ref. \cite{Thomas1} with
analytical model (table 1). Black points with error bars are the
experimental data. Green line is the coherent model. Blue line is
the model with loss of coherence and red line is the model with loss
of coherence and detector convolution. \label{Figure1}}
\end{figure}

We observe by red line of Fig. 4 that an analytical model including
loss of coherence as well as finite detector area is in full
agreement with existing experimental data. On the other hand, the
blue line shows that considering only loss of coherence there is an
agreement between the model and the experimental data but to obtain
a full agreement it is necessary to consider the detector
convolution. Green line shows that is not possible to adjust the
data by considering a completely coherent model. In the adjustment a
given value of partially coherent Gouy phase is necessary. The small
value found here is related with the set value of parameters used in
the experiment, specially the propagation times. Therefore,
different values of partially coherent Gouy phase can be obtained if
the experiment is realized with different set value of parameters.

\section{Conclusions}
We developed a theoretical model for the Poisson's spot problem by
using the Babinet principle. It was possible to include loss of
coherence and detector convolution in the observed intensity. Firstly,
we treated the coherent model and then we studied the effect of
the loss of coherence. Based in the previous definition to the Gouy
phase for partially coherent light waves (source incoherence) we
obtained an expression to the Gouy phase for partially coherent
matter waves (source incoherence + environment decoherence). We
observed that this phase influences the Poisson spot intensity. Therefore,
we have found a relationship between two old physical problems (Gouy
phase and Poisson spot). We obtained full agreement between our
results and existent experimental data. We observed that the Gouy
phase depends on the set value of parameters used in the Poisson
spot experiment. Thus, the Poisson spot experiment can be used to
measure the Gouy phase for partially coherent matter waves.

\vskip1.0cm
\begin{acknowledgments}
The authors would like to thank CNPq-Brazil for financial support.
I. G. da Paz thanks support from the program PROPESQ (UFPI/PI) under
grant number PROPESQ 23111.011083/2012-27.
\end{acknowledgments}

%\pagebreak

\end{document}